# Energy-Efficient Cooperative Caching in UAV Networks

Mohammad G. Khoshkholgh *Member, IEEE* and Victor C. M. Leung, *Fellow, IEEE*

*Abstract*—For an unmanned aerial vehicle (UAV) enabled network we investigate the energy-efficiency (EE) of joint caching and cooperative communication (Fog-RAN). Since UAVs are battery- and cache-limited, placing the popular contents in the caches and managing the energy expenditure of UAVs become crucial. We formulate the energy consumption of UAVs as an aggregate of communication/caching, hovering, and vertical displacement energies, and then devise an optimization problem for optimally assigning contents to caches and choosing the height of UAVs. Adopting tools from stochastic geometry, we also derive the EE in a numerically tractable form as a function of density, the radius of the cooperation zone, cache size, main communication/physical characteristics of UAVs, and influential environmental parameters. We develop two content placement strategies with low computational complexity. The conducted numerical results demonstrate that by adopting these algorithms one is able to improve EE by up to $800\%$ compared to common content placement schemes, e.g., the least-frequently used (LRU), the most-popular, and Hit-rate. Furthermore, while under LRU and Hit-rate schemes there is no benefit in vertically displacing UAVs, under our algorithms one is able to increase EE by at most $600\%$. Importantly, via our algorithms one can increase the size of cooperation zone in order to steadily increase EE, which is not the cases of LRU, the most-popular, and Hit-rate schemes. We finally observe that there is optimal values for density and cache-size of UAVs, granting maximum EE.

## I. Introduction

### A. Motivation

To provide "supper connectivity" in the terrestrial wireless communications the use of unmanned areal vehicles (UAVs), or drones, is proposed [1, 2]. Equipped with communication, processing, and storage capabilities, UAVs are able to operate as aerial (flying) base stations (BS) with adaptable mobility advantage that can circumvent the occurrence of severe blockages in cellular networks. This is because in comparison with ground communication the air-to-ground (A2G) communication more often experiences strong line-of-sight (LOS) links [3, 4]. On the other hand, when UAVs are endowed by sophisticated processing units they can be instrumental in assisting the ground UEs by on-demand offloading the heavy computation tasks in applications such as virtual reality and object recognition [4, 5]. UAVs are also influential for efficiently gathering a massive amount of data in IoT and machine-type applications in an energy efficient manner [2].

The UAV communication is in its infancy, and many challenges including optimal deployment, trajectory or path planning, A2G channel modeling, energy efficiency (EE), caching, and radio resource management are not well investigated [4]. These research problems have spurred substantial acclivities in academia and industries recently.

The authors of [3] obtained the optimal altitude of UAVs, providing maximum coverage. 3-D placement of UAVs in an on-demand UAV-enhanced cellular network was tackled in [6]. In [7], the authors proposed the join trajectory and resource allocation design. A practical mathematical tool is developed to assist coping with the sheer complexity of trajectory planning in a continues 3-D space, which is found sufficiently flexible for tackling many other pertinent design issues in UAV networks, see the references therein for related topics. The advantages of Fog computing in UAV communication networks to address computational limits of sensor devices in the Internet of Things applications are explored in [8]. Therein it is specified that the maneuverability of drones is crucial for effectively dealing with low computational capability of ground devices. Moreover, in [9] the trajectory optimization is focused for offloading computation-intensive tasks from devices to drones. Using stochastic geometry— a comprehensive tutorial of the subject along with related applications in wireless communication networks shall be found in [10]—the authors in [11] studied the spectrum sharing in drone communication networks and elaborated on suitable techniques for adjusting the density of drones with accordance to the required coverage probability on the ground.

However, to achieve seamless connectivity, the UAVs are obliged to frequently communicate with the high altitude platform or/and ground BSs. This operation heavily relies on the on-time communication between drones and terrestrial BSs via front-hauls to retrieve data from the core. As the current design of cellular networks has chiefly targeted the ground communications (which favors the down-tilted BSs) the high-speed, reliable communication between UAVs and the currently-deployed ground BSs is too difficult to be achieved [12]. In effect, since the ground-to-air (G2A) channel is LOS dominant one can expect that even very faraway ground BSs become potential interferers.

There are several plausible mechanisms to tackle this issue: 1) designing ground BSs for G2A communication by electronically/mechanically adjusting the antenna tilt of BSs. Combining this technique with massive MIMO and mmWave communications one is then able to potentially eradicate interference nearly entirely. This may not however stand as the best feasible solution for the near future, as it mandates significant reconfigurations of BSs that is both costly and time consuming. 2) Decreasing the frequency reuse for G2A communication along with aggressive interference cancellation mechanisms. As mentioned even very faraway ground BSs are able to pose significant interference, thus one requires to aggressively reduce frequency reuse. Also, the transceivers must employ efficient interference cancellation algorithms that require large processing capabilities and availability of on-time, accurate channel state information. 3) Caching the popular contents in the memory storage of UAVs. This scenario reduces the frequent communications between ground BSs and drones to occasional communications merely targeting the update of contents, the command and control procedures, and the like. Caching is therefore a feasible scenario at the moment as it imposes minimum network upgrades.

Nevertheless, caching systems are mainly studied for the ground communications. It is not obvious how to design caching in drone networks while exploiting unique characteristics of such networks and dealing with limited battery capacity of drones. This stands as the main motivation of this work. We are interested to spot unique features of caching in drone networks, to understand how different environments affect the caching performance, and to jointly design the content placement and mobility of drones.

*B. Literature Review*

The benefits of caching in traditional wireless networks has considerably explored, see, e.g., [13] for techniques and recent developments in caching in wireless networks. In general, in caching systems two core issues should be taken care of: *content placement*, which content should be placed in the cache of which BS, and *content delivery*, how to efficiently deliver contents upon their requests. In essence, it is desired to place contents to increase the hit rate—the probability that the content is find in a cache. An optimal randomized caching is proposed in [14] for small-cell networks, and further extended to $K$-tier heterogenous cellular networks (HetNets) [15, 16] and further to cooperative communications in [17–19]. Contrary to caching in the core, in wireless networks it is not optimal to merely cache the most popular contents due to coverage overlap of BSs. Furthermore, it is shown that by probabilistically caching contents at the small-cell BSs a considerable coverage performance is achievable while reducing the reliance on the back-haul communications.

The contributions made in [17–20] highlight the central role of physical-layer cooperation across adjacent BSs and effectiveness of randomized edge-caching in contemporary wireless networks. In [17] impact of a cluster-based co-operative caching along with orthogonal spectrum access, spectrum sharing, and interference cancellation is studied. It is shown that in caching systems there exists a tradeoff between communication diversity and caching diversity. In [19] a combination of the most-popular caching and probabilistic caching is suggested. Via analysis authors derive the percentage of the cache space that should be designated to the less popular contents. In 5G cellular networks, Fog-RAN endows the networking functionalities by introducing distributed edge computing and caching at the distributed remote antenna ports. The design of caching in Fog-RAN is investigated in [20]. Impact of BS height on the performance of cooperative caching in HetNets is also explored in [21]. Authors show that under COST 231 Hata path-loss model the average EE of the network reduces by increasing the height of BSs. Exploiting users preferences, the authors in [22] developed centralized and distributed content placement schemes. Results indicate that cooperative caching is instrumental in reducing the average download delay of the users. The authors of [23] developed a spatial signal alignment technique in Fog-RAN to efficiently mitigate interference at the victim receivers. Based on this novel cooperative scenario the remote heads that transmitting the same content simultaneously choose a specific class of modulation and coding enabling the end users to harness more received signal power and experience less interference. Work of [24] studied the EE for delivering scalable video services in Fog-RAN. The introduced solution segments the cooperation zone into two disjoint areas one responsible for caching/delivering the video contents with standard quality and the other responsible for caching/delivering the enhanced version of the video contents.

Owing to its many advantages, caching in UAV-enabled networks is also appealing and has received some attractions. In [25] distributed caching is advocated in UAV-assisted cellular networks. Simulation results suggest that proper interference alignment mechanism is crucial to secure the content delivery against eavesdropping. The authors of [26] used game theory along with machine learning to develop distributed resource-allocation algorithms based on the prediction of content request patterns. Furthermore, in [27], caching is exploited to defeat acute endurance problem, coming from the limited battery capacity of UAVs. UAVs are designed to optimally move so that they can cover the entire coverage area. The ground users are responsible to cache the contents for future delivery upon their requests. Furthermore, the authors of [28] discuss the probabilistic caching in UAV-enabled networks with the aid of stochastic geometry. It is shown that in high SNR regimes the probabilistic caching outperforms the most popular content placement strategy. However, in both [27] and [28] only a single-cell scenario is considered, i.e, the impact of interference among A2G links are ignored.

On the other hand, work of [29] demonstrates the impact of power control in UAV-assisted ultra-dense heterogenous cellular networks. The authors discuss that the power control is influential to improve the communication performance, caching performance, and energy transfer. Furthermore, via simulations it is shown that a UAV-assisted ultra-dense HetNet has higher performance compared to a stand-alone ultra-dense HetNet.

As seen, compared to the caching literature of terrestrial networks, the literature of caching in drone communications is minuscule and there are still many unresolved problems. Furthermore, in spite of its numerous advantages, the values of Fog-RAN for UAV communications has not yet fully discussed. So, we focus on UAV-enabled Fog-RAN and investigate energy-efficient caching and cooperative communications. Such an analysis is important and not trivial, recognizing the existence of various distinctive discrepancies between terrestrial and aerial communications. Relevant to our goal, the literature of terrestrial networks—except [21]—considers the simplistic standard path-loss model that overlooks the impact of LOS/non-LOS propagation. However, in UAV communications the path-loss model is far from the standard path-loss model of terrestrial networks and exhibits a probabilistic LOS/NLOS characteristic, depending on the altitude of drones as well as the environmental parameters of the target area, e.g., high-rise, sub-urban, and the like. We should also mention that [21] has not considered the content placement problem and only studied the content delivery phase. Furthermore, the main focus of study in [21] is spectral efficiency, and analyzing the EE is omitted. While in the terrestrial communications the main source of transmitter energy consumption is related to the communication and caching [30], in the UAV communication a substantial part of the energy consumption is associated with hovering and movement of drones. It is therefore crucial to correctly account for all main sources of energy in order to design sustainable UAV-enabled Fog-RAN.[1]

### C. Paper's Contributions

The main contributions of this paper can be summarized as follows:

- Considering energy consumption of communication/caching, hovering, and vertical displacement we formulate the total energy consumption of UAVs, and devise an optimization problem for optimally assigning contents to caches, subject to the cache size of UAVs, and choosing the height of UAVs.
- Adopting tools of stochastic geometry we derive EE performance in a numerically tractable form as a function of density, cooperation zone radius, cache size, main communication/physical characteristics of UAVs, and influential environmental parameters. Simulation results corroborate the analysis.
- We develop two new content placement strategies, namely, *recursive scaled hit rate (RSHR)* and *mixed popular-randomized caching (MPRC)*. Both schemes have very light computational burdens that is important in current caching systems with large catalog size.

RSHR is built upon a highly popular content placement strategy *Hit-rate*, see, e.g., [14, 17, 19, 32], but with a practical trick of recursively scaling hit rate probability of each content with the *soft-max* distribution of its EE performance. Although this appears to be a heuristic trick, it is capable of improving the EE remarkably. On the other hand, MPRC assigns a part of cache to the most poplar contents and the rest for the less popular contents. Via a simple greedy search—with the complexity in order of the cache size of UAVs—the algorithm balances these two parts in order to maximize the EE. In sprit, this algorithm is similar to the one developed in [17]. In comparison to [17], MPRC has much lower computational complexity and sweeps all the unpopular contents.

In this paper we also articulate two practically-appealing content placement scenarios in cooperative caching, leading to the most-popular content placement (MPCP) strategy: cache to minimize the interference in the cooperation zone and cache to maximize the capacity.

Extensive numerical studies demonstrate that RSHR improves the EE by up to 800% compared to common content placement schemes of the literature, e.g., least-frequently used (LRU), MPCP, and Hit-rate. More, while under LRU and Hit-rate schemes there is no benefit in vertically displacing UAVs, under RSHR (resp. MPRC) one is able to increase the EE by up to 600% (resp. 400%). Importantly, via our algorithms one can increase the size of cooperation zone in order to steadily increase the EE, which is not the cases of LRU, MPCP, and Hit-rate schemes. Last, we observe that there is the optimal density/cache-size of UAVs that maximizes the EE.

The rest of this paper is organized as follows. In Section II we discuss the system model using stochastic geometry. We further develop cooperative caching, formulate energy consumption of UAVs, and devise an optimization problem to address optimal content placement strategy and vertical displacement of UAVs. We analytically evaluate the EE in Section III. In Section IV we develop several content placement algorithms. In Section V we present numerical studies and evaluate the performance of our algorithms. This paper is finally concluded in Section VI.

## II. System Model, Network Model, and Cooperative Caching

In this section we discuss the channel model in UAV communications. The considered model is general enough and includes the small-scale fading, large-scale shadowing, and LOS/NLOS path-loss model. We further model the UAV network with the aid of stochastic geometry, introduce caching in UAV networks, discuss the cooperative edge-caching, and model the energy consumption of UAVs for delivering contents. We finally devise a proper optimization problem targeting the maximization of EE. A complete list of main parameters and notations can be found in Table I.

### A. Network Model

We adopt stochastic geometry to model the network. The location of UAVs (drones) in the 3-D space are accordingly drawn from a Homogenous Poisson point process (HPPP) $\Phi = \{(X_i, H) \in \mathbb{R}^3, i = 1, 2, \ldots\}$, where $X_i \in \mathbb{R}^2$ is the projection of the location of UAV $i$ on the 2-D plane and $H$ (km) is its altitude, which is the same across UAVs. The density of UAVs is $\lambda$, average number of UAVs per unit of area in km$^2$.

---

[1] In [31] we introduce the randomized caching in UAV-enabled Fog-RAN. Therein we show that cooperative communication is important for dealing with excessive LOS interference in drone communications. However a holistic account of energy consumption and the impact of caching on it has not explored. Furthermore, an accurate investigation of EE is missing in [31]. Last, in [31] we do not cover the development of computationally efficient algorithms for content placement and height of drones.

Consider the content library (catalog) $\mathcal{F} = \{f_1, f_2, \ldots, f_F\}$ with finite size $F = |\mathcal{F}| > 0$ in which $f_c$ is the $c$-th most popular file in the library. We interchangeably refer to a specific content either via $f_c$ or simply $c$. The popular contents are chosen and then sorted by their popularity in advance by adopting advanced big data analytics and machine learning. Let denote $a_c \in [0, 1]$ the popularity of content $c$. Here we assume that contents are indexed according to their popularity, i.e., $a_1 \leq a_2 \leq \ldots \leq a_F$.

The operation of UAV-enabled F-RAN is divided in two phases: *content placement* and *content delivery*. The former usually operates during off-the-peak traffic periods (early in the morning), locating the popular contents to the caches in the UAVs. Each UAV's cache size is $S$, which, in practice, can be much smaller than the catalog size $F$, i.e., $S \ll F$. We assume uncoded caching. Similar to [20, 21, 23, 30] we consider a cooperative communication scenario in which the requested content, $f_c$, is delivered to the UE by several adjacent UAVs with the aid of distributed beamforming, also known as non-coherent joint transmission, see, e.g., [33]. Since the content is cached at the UAVs, establishing a high-data rate backhaul links among the UAVs is not required, therefore advantages of coordinated multi-point (CoMP) communication can be exploited to improve the performance of the content delivery phase [17, 18, 33]. Assuming that the typical UE is located at the origin, all those UAVs within the circular cooperative zone of $X_{\text{cop}}$ (km) transmits $f_c$ to the typical user, if $f_c$ exists in their corresponding caches [20, 21, 30].

For cache-enabled terrestrial communications, various techniques to locate contents in the caches of small-cell BSs have been proposed, see, e.g., [14, 15, 34, 35] and references therein. Here, as [14–18], we consider randomized (probabilistic) content placement and work to extend it to UAV-enabled Fog-RAN. To this end, UAV $X_i$ caches $f_c$ with the probability $p_c \in [0, 1]$, where $\sum_{c=1}^{F} p_c = S$. Let further define $\mathcal{C} = \{\mathbf{p} : \sum_{c=1}^{F} p_c = S, p_c \in [0, 1]\}$ as a set of all feasible content placement probabilities. Therefore, the UAVs with $f_c$ in their cache belong to set $\Phi_c \subseteq \Phi$ and forms an HPPP with density $\lambda p_c$. Further, $\widetilde{\Phi}_c = \{X \in \Phi_c : \|X\| \leq X_{\text{cop}}\}$, is the set of UAVs able to engage in the cooperative transmission upon the request of file $f_c$.

### B. Channel Model

The channel between the UAVs and the users on the ground, known as *A2G channel*, is modeled as a combination of a large-scale path-loss attenuation, a large-scale shadowing, and a small-scale fading component [6, 12]. The A2G channel operates in LOS/NLOS modes [6], and the occurrence of LOS mode is shown to be dependent, among other things, on the drone's height, elevation angle, and the environment, e.g., dense urban or sparse rural. The probability that the channel between UAV $X_i$ and a receiver located at the origin, referred to as a *typical UE*, is an LOS channel specifies by a distance-dependent probability [2, 3, 6, 27]:

$$p_L(\|X_i\|) = \frac{1}{1 + \phi e^{-\psi\left(\frac{180}{\pi}\arctan\left(\frac{H}{\|X_i\|}\right) - \phi\right)}}, \quad (1)$$

where $\|X_i\|$ (km) is the 2-D Euclidian distance between the typical user and the UAV $X_i$, and $\phi$ and $\psi$ are the channel parameters capturing the traits of the underlying communication environment (refer to [3, 6] for nominal values). Increasing $H$ increases the probability of experiencing the LOS state, however with the cost of greater path-loss attenuation through attenuation function $L(\|X_i\|)$:

$$L(\|X_i\|) = \begin{cases} L_L(\|X_i\|) = \frac{K_L}{(\sqrt{H^2 + \|X_i\|})^{\alpha_L}} & \sim p_L(\|X_i\|), \\ L_N(\|X_i\|) = \frac{K_N}{(\sqrt{H^2 + \|X_i\|})^{\alpha_N}} & \sim p_N(\|X_i\|). \end{cases} \quad (2)$$

where $\alpha_L$ ($\alpha_N$) is the LOS (NLOS) path-loss exponent and $K_L$ ($K_N$) is the corresponding intercept constant.

For the fading fluctuations, we consider both small-scale power fading, $W_X$, and large-scale shadowing, $V_X$. The former is modeled using normalized (unit mean) Nakagami fading:

$$W_X = \begin{cases} W_X^L = \Gamma(\overline{W}_L, \frac{1}{\overline{W}_L}) & \sim p_L(X) \\ W_X^N = \Gamma(\overline{W}_N, \frac{1}{\overline{W}_N}) & \sim p_N(X), \end{cases} \quad (3)$$

where $\Gamma(a, b)$ is the Gamma distribution with parameters $a$ and $b$. Depending on the LOS/NLOS status of the communication channel between the UAV $X$ and the UE, the parameters $a, b$ will be different. It is reasonable to assume that $\overline{W}_L > \overline{W}_N$ as the fading is often more severe in NLOS channels. For the large-scale shadow-fading, we adopt a Log-normal model with the shadowing power gain $V_X = 10^{U_X/10}$:

$$U_X = \begin{cases} U_X^L \sim \mathcal{N}(\mu^L, \sigma_X^L) & \sim p_L(X) \\ U_X^N \sim \mathcal{N}(\mu^N, \sigma_X^N) & \sim p_N(X), \end{cases} \quad (4)$$

and $\mathcal{N}(\mu, \sigma)$ denotes a normal distribution with mean $\mu$ and variance $\sigma^2$, and $\sigma_X^{n_X}$ is given in [3]:

$$\sigma_X^{n_X} = a_{n_X} e^{-c_{n_X} \frac{180}{\pi}\arctan\left(\frac{H}{\|X\|}\right)}, \quad (5)$$

where $n_X \in \{L, N\}$, and $a_{n_X}$ and $c_{n_X}$ are channel parameters that depend on the communication environment. For several A2G communication scenarios values of path-loss model and shadowing parameters $a_{n_X}$ and $c_{n_X}$ are tabulated in [3].

Let's denote $P_r(X)$ as the received signal power from UAV $X$, i.e., $P_r(X) = L(\|X\|)V_X W_X$. As seen, $P_r(X)$ depends on the path-loss attenuation, large scale showing, and small-scale fading. Assuming $x = \|X\|$ is fixed, the average received signal power from UAV $X$, $\overline{P}_r(x) = \mathbb{E}[L(\|X\|)V_X W_X]$, can be evaluated as

$$\overline{P}_r(x) = \sum_{l_x \in \{L, N\}} p_{l_x}(x) L_{l_x}(x) \mathbb{E}[V_x^{l_x}] \mathbb{E}[W_x^{l_x}]$$

$$= \sum_{l_x \in \{L, N\}} p_{l_x}(x) L_{l_x}(x) e^{\left(\mu^{l_x} + \frac{(\sigma_x^{l_x})^2}{2}\right) \log 10}, \quad (6)$$

where we use the fact that fading (unit-mean) and shadowing are independent random variables.

TABLE I
LIST OF PARAMETERS AND NOTATIONS

| Symbol | Description |
| --- | --- |
| $\Phi = \{(X_i, H)\}$ | PPP with density $\lambda$ |
| $F$ | content library size |
| $S$ | cache size of UAVs |
| $a_c$ | popularity probability of file $c$ (the $c$th most popular content) |
| $X_{\text{cop}}$ | cooperation radius |
| $p_c$ | caching probability of file $c$ |
| $H_1$ | path-loss function between UAV $X$ and the origin |
| $n_X\{L, N\}$ | LOS/NLOS status of link UAV $X$ and origin |
| $W_X$ | Nakagami power fading in link UAV $X$ and origin |
| $V_X$ | large scale shadowing power gain in link UAV $X$ and origin |
| $\sigma_{V_X}$ | standard deviation of shadowing power gain $V_X$ |
| $P_r(X) = L(\|X\|)W_X V_X$ | received signal power from UAV $X$ at the origin |
| $\text{SINR}_c$ | SINR at the origin associated with content $c$ |
| $H_0$ | initial altitude of UAVs |
| $H_1$ | altitude of UAVs after vertical displacement |
| $E_{\text{tot}}$ | total consumed energy of each UAV |
| $E_{\text{com}}$ | communication energy expenditure |
| $E_{\text{hov}}$ | consumed energy for hovering |
| $\overline{T}_t$ | communication window |
| $v_v$ | vertical displacement speed of UAVs |
| $P$ | Transmission power of UAVs |
| $P_h(H)$ | hovering power |
| $\zeta$ | direction of displacement of UAVs |
| $P_v(\zeta)$ | displacement power |
| $P_s$ | circuit power |
| $P_p$ | caching power |

## C. SINR Formulation

Our main goal is to investigate the energy efficiency (EE) of edge-caching in cooperative UAV communications.

To formulate the EE of the introduced cooperative caching we first formulate the data rate of the typical UE, $R_c$, to account for impacts of the cache size, altitude of UAVs, the size of cooperation zone, and other pertinent system parameters. For the typical UE requesting $f_c$, the received power of the information-bearing signal is $\sum_{X \in \tilde{\Phi}_c} P_r(X)$, where, as it is specified above, $\tilde{\Phi}_c$ is the set of UAVs with $f_c$ in their cache which are hovering over the cooperation zone. The post-processed received signal-to-interference and noise (SINR) is then written as

$$\text{SINR}_c = \frac{\sum_{X \in \tilde{\Phi}_c} P_r(X)}{\sum_{Z \in \Phi \setminus \Phi_c} P_r(Z) + \sum_{Y \in \Phi_c \setminus \tilde{\Phi}_c} P_r(Y) + \frac{\sigma^2}{P}}, \quad (7)$$

where $\sigma^2$ is the power of AWGN and $P$ Watts is the transmission power of UAVs. In (7), the first term in the denominator represents the received interference from the active UAVs without $f_c$ in their cache while the second term captures the received interference from those UAVs with $f_c$ in their cache located outside of the cooperation zone. Denoting the spectrum bandwidth by $B$ Hz, the best achievable transmission data rate (bit/sec) of delivering $f_c$ is effectively obtained from Shannon formula

$$R_c(H) = B \log \left(1 + \text{SINR}_c\right). \quad (8)$$

## D. Modeling Energy Consumption of UAVs

In UAV-enabled communications, energy-efficiency (EE) is important as drones have a limited energy storage on board [1, 30]. We thus investigate the EE of the edge caching.

We assume UAVs are initially located at altitude $H_0$, which is assumed to be fixed and given. In general, $H_0$ could be optimized, for instance, to maximize the throughput or minimize delay for communication between UAVs and the ground BSs or UAVs and the top layer satellite system/high-altitude platforms (HAP). Such a communication channel may take place, for instance, through designated high-speed macrowave/mmWave backhaul/fronthaul channels supporting UAVs for command-and-control communications and/or updating/replacing contents at the caches. One may also optimize $H_0$ for wireless power transfer, required for recharging the battery of UAVs.

Although the initial altitude of UAVs are fixed to $H_0$, UAVs are permitted to be *vertically displaced* in order to (possibly) improve the EE performance of the network. Assume UAVs are displaced to altitude $H_1$. Note that the projection of UAVs locations on the ground is not affected, thus the chance of collision among UAVs is quite remote, given than their initial locations were drawn from PPP.[2] The total energy expenditure of each UAV, $E_{\text{tot}}$, is basically comprised of three main parts: 1) the required energy for communication, memory, and processing, $E_{\text{com}}$, 2) the required energy to keep the UAVs hovering during communication window, $E_{\text{hov}}$, and 3) the required energy for displacing the UAVs from altitude $H_0$ to altitude $H_1$, $E_{\text{dis}}$. We now model all these components.

*1) Communication Energy:* Communication energy is related to the transmission and circuit power. Caching also consumes energy, where the amount of the required energy generally depends on the cache size and its underlying memory technology, e.g., solid state disk (SSD), dynamic random access memory (DRAM) [35]. Let the required power by a unit of cache be $P_p$ Watts. Then, a UAV consumes $SP_p$ for caching. There are also circuit power consumptions, where the former, denoted by $P_s$, is often constant. Assuming UAVs are vertically displaced by speed $v_v$ m/sec (in practice $v_v$ can be up to 460 km/h [4]), accounting for the displacement time, which is equal to $\frac{|H_1 - H_0|}{v_v}$, only $\left(\overline{T}_t - \frac{|H_1 - H_0|}{v_v}\right)^+$ seconds of the available communication window $\overline{T}_t$ remains available for the actual content delivery. Note that in reality the time duration that UAVs are operating can be substantially larger than $\overline{T}_t$. Our main concern is to maximize the energy consumption per each communication window. Therefore, $E_{\text{com}}$ is formulated as

$$E_{\text{com}} = \left(\overline{T}_t - \frac{|H_1 - H_0|}{v_v}\right)^+ P + (P_s + SP_p). \quad (9)$$

The underlying assumption here is that during movement control the communication is aborted, which is the case of the practice [4].

*2) Hovering Energy:* Denote $P_h(H)$ the hovering power of each UAV at the altitude $H$. In general, $P_h(H)$ depends on the several characteristics of the UAV, such as weight and motor, as well as atmospheric conditions including air density $\rho(H)$. The air density is usually given for the nominal sea level temperature 15 $^oC$ and height, which is $\rho_0 \approx 1.225$kg/m$^3$. However, it should be adjusted based on the altitude through $\rho(H) \approx \rho_0 e^{-c_\rho H}$ where $c_\rho \approx 0.118$(km)$^{-1}$ is a constant depending on environmental conditions [36]. Accounting for the change of air density with height, for a rotary-wing aircraft in hovering status one can adopt the calculations in [37] to obtain the corresponding power consumption as the following

$$P_h(H) = c_1 \rho(H) + \frac{c_2 M^{1.5}}{\sqrt{\rho(H)\pi d^2}}, \tag{10}$$

where $M$ (kg) is the weight of UAV$^3$, $d$ (m) is the rotor disk radius, $c_2 = 1.1$, and $c_1$ is a function of UAV's rotor characteristics, which can be approximated as $c_1 = 1.91$ [37]. Using (10), the hovering energy is then formulated as

$$E_{\text{hov}} = \left(\overline{T}_t - \frac{|H_1 - H_0|}{v_v}\right)^+ P_h(H). \tag{11}$$

---

[2]The inclusion of horizontal mobility or a combination of horizontal and vertical mobility of drones in a stochastic geometry analysis of the network is inherently complex and to some levels intractable. This is due to the fact that horizontal mobility induces new levels of correlation in the interference across time slots, which its correct characterization requires information regarding the trajectory of all drones during communication window. Since such information may not be available, mathematical models eventually seem necessary to model the mobility of drones. Such mathematical models should adhere to the actual trajectory solutions developed in the literature.

*3) Displacement Energy:* To formulate $E_{\text{dis}}$ we should account for the direction—ascending or descending—of displacement too, as the consumed power to support the displacement is basically different in ascending and descending directions. Denote parameter $\zeta \in \{\text{up}, \text{down}\}$ to specify the direction of displacement where $\zeta = \text{up}$ (resp. $\zeta = \text{down}$) stands for the case that UAVs are ascending (resp. descending). Accordingly, $P_v(\zeta)$, the required power to displace each UAV, can be obtained from [2]$^4$

$$P_v(\zeta) = \begin{cases} \frac{M}{2} v_v + \frac{M}{2}\sqrt{v_v^2 + \frac{2M}{\rho \pi d^2}} & \zeta = \text{up}, \\ \frac{M}{2} v_v - \frac{M}{2}\sqrt{v_v^2 - \frac{2M}{\rho \pi d^2}} & \zeta = \text{down}. \end{cases} \tag{12}$$

Note that when the UAV is descending we need to guarantee $v_v \geq \sqrt{\frac{2M}{\rho \pi d^2}}$. As a result, the required energy to support the displacement over distance $|H_1 - H_0|$ is

$$E_{\text{dis}} = \frac{|H_1 - H_0|}{v_v} P_v(\zeta). \tag{13}$$

---

$^3$Note that $M$ can be a function of cache capacity. For example a 10 TB Seagate 3.5 HDD (Helium) weights around 0.655kg. Generally it is hard to model the weight of SSD/HDD as a function of its storage capacity, and it varies substantially depending on the technology, brand, and several other physical/functional characteristics. Therefore, in this paper we assume that $M$ does not increase with $S$. This is an acceptable assumption given that the installed cache has enough storage capacity to begin with, say, 10 TB. Thus it is possible to accommodate as many contents (with reasonable size, e.g., less than 1 GB) as necessary.

$^4$Various mathematical models are developed in order to precisely formulae the power consumption of drones depending on weight, type of drone (fixed-wing versus rotary-wing), blade profile of the drone (blade chord, number of blades, blade power profile, and the frontal area of drone, and the like), induced hovering power with accordance the tilt angle of the drone (which depends on rotor disk radius, angular velocity of drone, and the weight of drone), ascending/descending direction of drone, and the like. see, e.g., [38–41]. Incorporation of accurate model of power consumption when the subject of the analysis is the wireless communication has its own complexity. The related literature, see, e.g., [2], [37], has made some progress to capture the most relevant aspects of this theory for the purpose of wireless communication networking. We in this paper follows this approach.

*4) Total Energy Consumption:* Now, summing up (9), (11), and (13), the total energy consumption of each UAV is formulated as $^5$

$$E_{\text{tot}} = \left(\overline{T}_t - \frac{|H_1 - H_0|}{v_v}\right)^+ (P + P_h(H_1))$$
$$+ \frac{|H_1 - H_0|}{v_v} P_v(\zeta) + (P_s + SP_p). \tag{14}$$

### E. Problem Formulation

Our goal is to place contents in the caches (withe limited capacity $S$) of UAVs in order to maximize the energy efficiency of content delivery phase. Furthermore, we desire to properly displace UAVs from initial altitude $H_0$ to $H_1$ for possible improvement of the EE. Note that since displacement costs power and reduces the communication window, depending on the speed, $v_v$ and the displacement distance, $|H_0 - H_1|$, the displacement may not necessarily become recommendable. For each content $c$, using (8) and (14), the EE is:

$$\eta_c(H_1, \zeta) = \frac{\left(\overline{T}_t - \frac{|H_1 - H_0|}{v_v}\right)^+}{E_{\text{tot}}} \gamma_c(H_1, p_c). \tag{15}$$

where for given altitude $H$ and content probability $p_c$, $\gamma_c(H, p_c)$ is related to the capacity given in (8) via

$$\gamma_c(H, p_c) = \mathbb{E}\left[\mathbb{1}_{|\tilde{\Phi}_c| > 0} \frac{R_c(H)}{|\tilde{\Phi}_c|}\right], \tag{16}$$

where $\mathbb{E}$ represents expectation over any randomness including position of the UAVs, path-loss attenuation, small-scale fading, shadowing, and content placement. In formulating EE we note that 1) $R_c(H)$ is the collective received data rate from $|\tilde{\Phi}_c|$ UAVs in the cooperation zone; 2) the total consumed energy in the cooperation zone is $E_{\text{tot}}|\tilde{\Phi}_c|$. EE in (15) therefore quantifies the total energy efficiency of UAVs in the cooperation zone for the delivery of content $c$. Using (15) and the law of total probability, the EE performance of the network is expressed as $\eta(H_1, \zeta) = \sum_c a_c \eta_c(H_1, \zeta)$. Accordingly, the joined content placement and UAV displacement problem is formulated through the following optimization problem:

$$\mathcal{P} : \eta^* = \max_{\boldsymbol{p} \in \mathcal{C}} \eta(H_1, \zeta) \tag{17}$$

$$s.t. \quad \zeta \in \{\text{up}, \text{down}\}, H_1 \in [H_{\min}, H_{\max}].$$

In this optimization problem $H_{\min}$ (resp. $H_{\max}$) stands for the minimum (resp. maximum) permissible altitudes, which

---

$^5$Beside the mathematical approaches of [38–41], another line of work focuses on developing a statistical model of energy consumption via measurements and curve fitting. For example, in [41] for particular brands "3D solo" and "DJI Matrice 100" the authors derive simple formulas that relate the weight and other physical characteristics, wind condition, and the mobility profile of the drones to the energy consumption of the drones. The author then showcase the value of the models by using them for the trajectory optimization. While such an approach has its own merits, it requires the existence of reliable models and sufficient measurements.

are usually enforced from regulatory bodies [4].[6]

Optimization problem $\mathcal{P}$ is substantially complex and may impose considerable computational cost, especially when the size of the content library $F$ is too large. On the other hand, as we also see from Proposition 1 in the next section, $\eta(H_1, \zeta)$ is not analytically amenable for differentiation with respect to content placement probability $p_c$ and altitude $H_1$. We therefore develop an heuristic solution. We first fix the direction, and partition the interval $[H_{\min}, H_{\max}]$ into $N_{\max} \gg 1$ non-overlapping intervals $[H_{\min}, H_{\max}] = \bigcup_l [\underline{H}_l, \overline{H}_l]$ where $\underline{H}_0 = H_{\min}$ and $\overline{H}_{N_{\max}} = H_{\max}$. We then set the altitude as $H_l = \frac{\overline{H}_l + \underline{H}_l}{2}$ for each interval $[\underline{H}_l, \overline{H}_l]$, and solve the associated cache placement problem:

$$\mathcal{P}_{\text{cache}} : \tilde{\eta}(H_l, \zeta) = \max_{\boldsymbol{p} \in \mathcal{C}} \sum_{c=1}^{F} a_c \eta_c(H_l, \zeta). \quad (18)$$

In Section IV we develop algorithms for solving $\mathcal{P}_{\text{cache}}$. Note that for ascending (resp. descending) direction the optimization problem $\mathcal{P}_{\text{cache}}$ needs to be solved $\lceil N_{\max} \frac{H_0}{H_{\max}} \rceil$ times (resp. $\lceil N_{\max} \frac{H_{\min}}{H_0} \rceil$ times). Assuming that the complexity of solving the optimization problem $\mathcal{P}_{\text{cache}}$ is $\mathcal{O}_{\text{cache}}$, the complexity of optimization problem (17) is therefore $(\lceil N_{\max} \frac{H_0}{H_{\max}} \rceil + \lceil N_{\max} \frac{H_{\min}}{H_0} \rceil) \mathcal{O}_{\text{cache}} \approx N_{\max} \mathcal{O}_{\text{cache}}$.

## III. PERFORMANCE ANALYSIS

To solve optimization problem (17) we require to obtain the EE as a function of main system parameters including the content placement probabilities, altitude of UAVs, the size of the cooperation zone, and the like. Regarding (15), to evaluate EE we need to obtain $\gamma_c(H, p_c)$, which is accomplished in the following proposition.

*Proposition 1:* Let define function

$$\mathcal{L}_1(x, v) = \sum_{l_x \in \{L, N\}} \frac{p_{l_x}(x)}{\sqrt{(2\pi \sigma_x^{l_x})^2}} \int_0^{\infty} \frac{e^{-\frac{(u - \mu^{l_x})^2}{2(\sigma_x^{l_x})^2}}}{\left(1 + v \frac{L_{l_x}(x) 10^{u/10}}{\overline{W}_{l_x}}\right)^{\overline{W}_{l_x}}} du.$$

Thus, for the considered cooperative communications, when content $c$ is requested $\gamma_c(H, p_c)$ is obtained from

$$\gamma_c(H, p_c) = \frac{2\pi \lambda p_c}{e^{\Theta_{\text{cop}} p_c}} \int_0^{\infty} \Omega(v, p_c) \int_0^{X_{\text{cop}}} \left( \frac{e^{\pi \lambda p_c (X_{\text{cop}}^2 - x^2)} - 1}{\pi \lambda p_c (X_{\text{cop}}^2 - x^2)} \right)$$

---

[6]**It is quite straightforward to extend the analysis of this paper to heterogenous UAV networks. Consider several independent tiers of UAVs each of which with specific cache capability (i.e., the cache size), cache strategy (e.g., caching probability), drone technology (e.g., weight, memory, communication power, and the like), communication requirements, e.g., SINR requirements and cooperation zone, and the density of drones. The problem is then to specify the altitude range of each tier (e.g., very low altitude, low altitude, medium altitude, high altitude, and very high altitude) along with caching parameters to maximize the energy efficiency. On the other hand since drone communication is very vulnerable to excessive LOS interference, besides altitude and caching strategy other issues such as interference management and spectrum allocation should be incorporated in the problem formulation. For example, one may decide to designate spectrum exclusively to each tier where the optimal bandwidths can be specified based on the QoE requirements of the end users, height as well as other characteristics of the drones.**

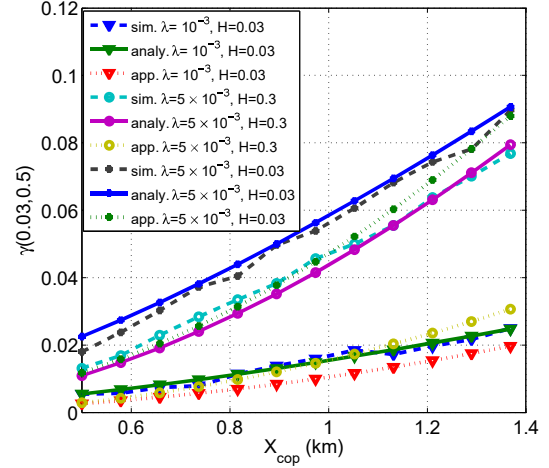

Fig. 1. $\gamma_c(H, p_c)$ versus $X_{\text{cop}}$ in the urban environment, when $F = 2$, $S = 1$, $p_c = 0.5$ and $H = 0.03$ km.

$$- \mathcal{L}_1(x, v) \frac{e^{\pi \lambda p_c \Psi(x, v)} - 1}{\pi \lambda p_c \Psi(x, v)} \Big) x \, dx \, dv \quad (19)$$

where $\Theta_{\text{cop}} = \pi \lambda X_{\text{cop}}^2$, $\Psi(x, v) = \int_x^{X_{\text{cop}}} 2y \mathcal{L}_1(y, v) dy$ and $\Omega(v, p_c) = \frac{e^{-v \frac{\sigma_r^2}{P}}}{v} \mathcal{L}_{I_c}(v)$, in which $\mathcal{L}_{I_c}(v)$ is the Laplace transform of interference:

$$e^{-2\pi \lambda (1 - p_c) \int_0^{\infty} z(1 - \mathcal{L}_1(z, v)) dz} e^{-2\pi p_c \lambda \int_0^{X_{\text{cop}}} z(1 - \mathcal{L}_1(z, v)) dz}.$$

*Proof:* See Appendix-A. $\square$

We further derive an approximate of $\gamma_c(H, p_c)$ in the following Proposition:

*Proposition 2:* Assume content $c$ is requested, $\gamma_c(H, p_c)$ can be approximated via

$$\gamma_c(H, p_c) \approx \sum_{n_c=1}^{\infty} \frac{(\Theta_{\text{cop}} p_c)^{n_c}}{n_c! n_c e^{\Theta_{\text{cop}} p_c}} \log \left( 1 + 2\pi \lambda p_c \int_0^{X_{\text{cop}}} \frac{x}{e^{\pi \lambda p_c x^2}} \big(}$$

$$\overline{P}_r(x) + (n_c - 1) \overline{P}_r(0.5x + 0.5X_{\text{cop}})\big) dx \int_0^{\infty} \frac{\mathcal{L}_{I_c}(v)}{e^{\frac{\sigma_r^2}{P} v}} dv \Big),$$

where $\mathcal{L}_{I_c}(v)$ is given in Proposition1, and $\overline{P}_r(x)$ is the average received power from UAV $X_i$ with $x = \|X_i\|$ and is given in (6).

*Proof:* See Appendix-B. $\square$

Compared to Proposition 1, the numerical complexity of Proposition 2 is substantially lower. Nevertheless, the cost of this numerical complexity reduction is the accuracy. This issue is explicitly seen in Fig. 1 and Fig. 2, where we depict $\gamma_c(H, p_c)$ versus $X_{\text{cop}}$ for sub-urban and high-rise environments, respectively. As also seen from these illustrations, Proposition 1 is accurate, we thus use it for solving the optimization problem (17).

On the other hand, from these illustrations we observe that in general increasing the cooperation zone radius improves $\gamma_c(H, p_c)$ since the chance of finding the content in the cooperation zone increases and the received signal power grows. We further note that by increasing the altitude $H$, $\gamma_c(H, p_c)$ reduces. This is perhaps due to the growth of free-space path-loss attenuation, which causes the received signal power to decline substantially. On the other hand, the accumulate interference can grow by increasing the altitude, since the LOS interfering signals become more dominant. This issue can be also substantiated by comparing Fig. 1 and Fig. 2: In high-rise environment $\gamma_c(H, p_c)$ is larger than that of the sub-urban environment. In effect, the LOS probability in sub-urban areas is much higher compared to the high-rise case.

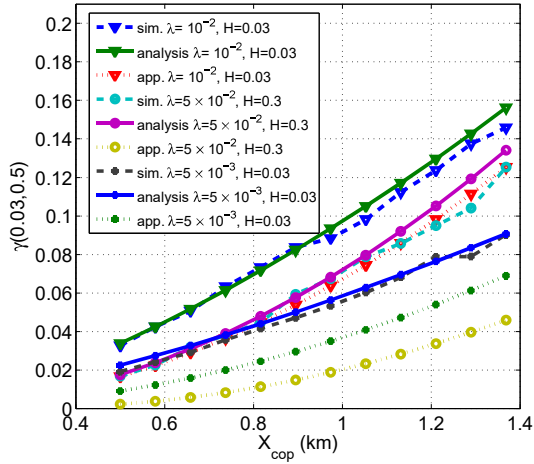

Fig. 2. $\gamma_c(H, p_c)$ versus $X_{\text{cop}}$ in the high-rise environment, when $F = 2$, $S = 1$, $p_c = 0.5$ and $H = 0.03$ km.

## IV. ALGORITHMS FOR ENERGY-EFFICIENT CONTENT PLACEMENT

As we discussed in Section II-E to solve optimization problem $\mathcal{P}$ we require to solve content placement problem $\mathcal{P}_{\text{cache}}$, roughly $N_{\max}$ times. This optimization problem is hard to tackle as the complexity of deriving the differentiation of the objective function with respect to content placement probabilities $\{p_c\}$ is not tractable. More, $\mathcal{P}_{\text{cache}}$ may not be a convex optimization problem. In effect, for practical scenarios that $F$ is very large (up to thousands) the off-the-shelf available numerical solutions, e.g., interior point method or ant colony algorithm, are deemed to pose impractically large computational complexity. Here we show how one can upgrade some available caching algorithms of the literature (mainly developed for the terrestrial networks with different objective function that is considered in this paper) in order to make them more numerically efficient.

### A. Most Popular Content Placement (MPCP) Strategy

Under MPCP strategy the most popular contents are located in all the cashes, $\mathcal{S}_X = [1, S] \ \forall X$. In content-centric systems (Internet caching), MPCP strategy is desirable. In wireless networks, it enhances hit rate when BSs have *non-overlapping service area* [34], which is not conceivably attainable in reality as several BSs can simultaneously provide coverage [30]. In the following we demonstrate two viewpoints in designing content placement in cooperative caching that lead to MPCP. Furthermore, in Section V we see that MPCP can attain a good portion of more sophisticated content placement schemes in some relevant scenarios. Acknowledging that MPCP is very simple to implement, this is a practically significant observation.

*1) Cache to Minimize In-Cluster Interference:* Assume file $f_c$ is requested. The activity of UAVs $X \in \bigcup_{c' \neq c} \Phi_{c'}$ is basically the main source of interference at the cooperation zone, reducing the achievable capacity. Since EE can increase by the growth of achievable capacity, one may then place contents to minimize the in-cluster interference. One is then recommended to monopolize the cooperation zone merely to file $f_c$, implying that 1) minimizing the effect of interference by enforcing $\bigcup_{c' \neq c} \Phi_{c'} \bigcap \mathcal{B}_O(\|X_{\text{cop}}\|) = \varnothing$ where $\mathcal{B}_O(r)$ is a disk with radius $r$ centered at origin, and 2) ensuring that the typical UE receives the requested file by enforcing $\tilde{\Phi}_c \neq \varnothing$. Recall that $\Theta_{\text{cop}} = \pi \lambda X_{\text{cop}}^2$. Thus, since contents are probabilistically located at the caches, the former event has the probability $\prod_{c' \neq c} e^{-\Theta_{\text{cop}} p_{c'}}$ and the latter event happens with probability $(1 - e^{-\Theta_{\text{cop}} p_c})$. Using these two probabilities, the objective function can be formulated as

$$\sum_c a_c (1 - e^{-\Theta_{\text{cop}} p_c}) \prod_{c' \neq c} e^{-\Theta_{\text{cop}} p_{c'}}$$
$$= e^{-\Theta_{\text{cop}} S} \sum_c a_c (e^{\Theta_{\text{cop}} p_c} - 1),$$

because $\sum_c p_c = S$. Now to specify the caching probabilities we solve the optimization problem

$$\max_{\boldsymbol{p} \in \mathcal{C}} \sum_c a_c (e^{\Theta_{\text{cop}} p_c} - 1).$$

The objective function of this optimization problem is convex. Therefore, the optimal point coincides with one of the boundary points. Since the objective function is monotonically increasing of caching probabilities the boundary point rendering the maximization of the objective function is therefore $p_c^* = 1_{c \leq S}$, thus $\mathcal{S}_X = [1, S]$.

*2) Cache to Maximize the Virtual Capacity:* In above approach we construct the scenario that (indirectly) leads to improvement of capacity (but not necessarily EE), by minimizing interference. Here we directly target the maximization of capacity. Nevertheless this problem becomes substantially complex as, similar to the case of EE in Proposition 1, the evaluation of capacity is numerically troublesome. We therefore promote the use of an estimate of the achievable capacity that we call *virtual capacity*. Assuming file $f_c$ is requested the virtual capacity is formulated as

$$R_c^{\text{vir-cap}}(p_c) = \log\left(1 + \frac{p_c \chi^{\text{sig}}}{\overline{I}(p_c)}\right),$$

where $\chi_c^{\text{sig}}(p_c)$ is the expected value of the information-bearing signals, i.e., $\chi_c^{\text{sig}}(p_c) = \mathbb{E} \sum_{X \in \tilde{\Phi}_c} P_r(X)$, and $\overline{I}(p_c)$ is the expected interference plus noise i.e.,

$$\overline{I}(p_c) = \mathbb{E} \sum_{Z \in \Phi \setminus \Phi_c} P_r(Z) + \mathbb{E} \sum_{Y \in \Phi_c \setminus \tilde{\Phi}_c} P_r(Y) + \sigma^2.$$

Note that the virtual capacity is not neither an upper-bound nor a lower-bound on the actual achievable capacity. However, it provides a rough estimation of the expected capacity when there are many uncertainties.

Using Campbell's Theorem [10] we can show that $\chi^{\text{sig}} = 2\pi\lambda \int_0^{X_{\text{cop}}} x\overline{P}_r(x)dx$, where $\overline{P}_r(x)$ is the average received signal power from the UAV located at $x$ obtained in (6). Furthermore, $\overline{I}(p_c)$ is calculated as

$$\overline{I}(p_c) = ((1-p_c)B + p_cC + \sigma^2)^{-1},$$

in which $C = 2\pi\lambda \int_{X_{\text{cop}}}^{\infty} x\overline{P}_r(x)dx$ and $B = 2\pi\lambda \int_0^{\infty} x\overline{P}_r(x)dx = \chi_c^{\text{sig}} + C$. Therefore, we have

$$\overline{I}(p_c) = (\chi^{\text{sig}} + C - p_c\chi^{\text{sig}} + \sigma^2)^{-1},$$

which, in turn, implies that

$$R_c^{\text{vir-cap}}(p_c) = \log(1 + \frac{p_c}{\Theta_{\text{vir-cap}} - p_c}),$$

in which $\Theta_{\text{vir-cap}} = 1 + \frac{C+\sigma^2}{\chi^{\text{sig}}}$. We now construct the optimization problem

$$\max_{p\in\mathcal{C}}\sum_{c=1}^{F} a_c \log(1 + \frac{p_c}{\Theta_{\text{vir-cap}} - p_c}),$$

to obtain $p$. Here the objective function is convex, thus the optimal solution must happen in boundary points. We also note that function $\log(1 + \frac{p_c}{\Theta_{\text{vir-cap}} - p_c})$ is increasing with respect to variable $p_c$. This implies that caching the $S$ most popular contents is the optimal solution, i.e., $\mathcal{S}_X = [1, S]$.

### B. Mixed Popular-Randomized Caching (MPRC) Strategy

As elaborated on in Section IV-A the MPCP targets the maximization of (virtual) capacity or the minimization of in-cluster interference. However, neither of these necessarily leads to the maximization of energy efficiency, which is the main goal of problem (18). Here propose a mixed popular-randomized caching (MPRC) scheme, where part of the cache space of each UAV is reserved for caching the most popular content, while the remaining is used for caching the less popular contents.

We divide each cache into two disjoint parts where the first part contains the most popular contents and the second part the less popular ones. Let $0 \leq S_{\text{pop}} \leq S$ be the size of caches assigned to the $S_{\text{pop}}$-most popular contents, which is a design parameter. The rest of cache space is assigned for $S_{\text{rnd}} = S - S_{\text{pop}}$ less popular contents. To fill it, each UAV randomly draws an index $c \in [S_{\text{pop}} + 1, F - S_{\text{rnd}} + 1]$, with probability $\frac{1}{F - S_{\text{pop}} - S_{\text{rnd}} + 1} = \frac{1}{F - S + 1}$ and caches contents with indices in $[c, c + S_{\text{rnd}}]$. Thus, $p_c$ is written as a function of $S_{\text{pop}}$ via

$$p_c = \mathbb{P}\{c \in \mathcal{S}_{x_i}\} = 1_{c \leq S_{\text{pop}}} + \sum_{m=S_{\text{pop}}+1}^{F-S_{\text{rnd}}+1} \frac{1_{m \leq c \leq m + S_{\text{uni}}}}{F - S + 1}$$

$$= \begin{cases} 1 & 1 \leq c \leq S_{\text{pop}}, \\ \frac{c - S_{\text{pop}}}{F - S + 1} & S_{\text{pop}} + 1 \leq c \leq S, \\ \frac{S_{\text{rnd}}}{F - S + 1} & S < c \leq F - S_{\text{rnd}} + 1, \\ \frac{F - c + 1}{F - S + 1} & c > F - S_{\text{rnd}} + 1. \end{cases} \quad (20)$$

Consequently, the approximated content placement problem can be written as

$$\max_{0 \leq S_{\text{pop}} \leq S}\sum_{c=1}^{F} a_c \eta_c(H_0),$$

which only requires a greedy search over $S_{\text{pop}}$.

*Remark 1:* MPRC is similar to the the algorithm developed in [17]. In comparison to [17], here we let the algorithm sweeps all the unpopular contents. On the other hand, algorithm of [17] has much higher computational cost compared to MPRC as it relies upon optimizing the performance over the design parameter that is a continuous variable in interval $[0, 1]$. However, the complexity of MPRC grows at most by $S$. ∎

### C. Recursive Scaled Hit Rate (RSHR) Algorithm

A popular randomized caching strategy is hit-rate maximization [14, 17, 19], whereby contents are placed in the caches so that the *hit rate*, $1 - e^{-\Theta_{\text{cop}} p_c}$, is maximized, i.e.,

$$\max_{p\in\mathcal{C}}\sum_c a_c(1 - e^{-\Theta_{\text{cop}} p_c}).$$

Undoubtedly, this algorithm may not maximize the EE. To tackle this issue we here introduce a modification to this algorithm in order to make it a proper option for maximizing EE. (Although the solution is solely tailored for maximization of EE, our solution is general enough and can be applied for any other objective functions straightforwardly.) Our approach is to recursively scale the hit-rate of content $c$ with accordance to the resulted EE of the content. Put in another word, in each iteration $t > 1$ we initially construct scales $\{b_c[t-1]\}$ via

$$b_c[t-1] = \frac{e^{\eta_c(H_0;t-1)}}{\sum_{c'=1}^{F} e^{\eta_{c'}(H_0;t-1)}}$$

using the (calculated) probabilities $p_c(t-1)$. It is readily to confirm that $b_c[t-1] \in [0, 1]$ and $\sum_c b_c[t-1] = 1$. On the other hand, we can see that $b_c[t-1] \geq b_{c'}[t-1]$ if $\eta_c(H_0; t-1) > \eta_{c'}(H_0; t-1)$. The scales form soft-max distribution with the score functions driven from the EE performance of contents.

Therefore, by upgrading the hit-rate in each iteration $t$ through $b_c[t-1]\left(1 - e^{-\Theta_{\text{cop}} p_c}\right)$ we promote contents that are more energy efficient. Assuming $b_c[1] = 1 \ \forall c$, in each iteration $t$ we therefore solve the optimization problem

$$\max_{p\in\mathcal{C}}\sum_{c=1}^{F} a_c b_c[t-1]\left(1 - e^{-\Theta_{\text{cop}} p_c}\right),$$

to derive $p_c^*(t)$ according to the algorithm developed in [14, 19]. We repeat the iterations until the objective function becomes stable. Our numerical studies (not included in the following) confirm that this algorithm converges in less than ten iterations.

## V. Numerical Study and Performance Evaluation

We now evaluate the impact of various system parameters on the EE. As the common practice of the literature, we model the popularity using Zipf distribution [14, 17, 19, 32]. The probability that $f_m$ is requested is $a_m = m^{-\kappa} (\sum_{c=1}^{F} c^{-\kappa})^{-1}$, where $0 \leq \kappa \leq 2$ is the skewness of the distribution referred to as *popularity exponent*. For $\kappa \to 0$, all contents become equally popular, i.e., uniform distribution. For simplicity, as in [14, 32], we assume the equal size files in $\mathcal{F}$. We compare the performance of MPCP, RSHR, MPRC, and commonly considered cases of LRU and Hit-rate [14]. The parameters of the A2G channel can be found in [3]. We also set $\alpha_L = 2.09$, $\alpha_N = 4$, $\overline{W}_L = 10$, $\overline{W}_N = 2$, $M = 10.2$ kg, $P = 1$ W, $P_p = 1$ μW, $P_s = 0.01$ mW, $T_t = 200$ sec, $d = 0.5$ m, $H_{\min} = 50$ m, and $H_{\max} = 400$ m. When $H_0$ is fixed, we set to be equal to $H_0 = 200$ m.

*1) Impact of Popularity Exponent $\kappa$:* Fig. 3 shows the impact of popularity exponent on the EE. As seen, the higher is the value of $\kappa$, the larger is the EE. We further observe that 1) in the high-rise environment the EE is much higher than that of the sub-urban environment. In effect, in high-rise building environment, the UAV signals will likely experience NLOS mode and the LOS signal could be blocked by the high-rise buildings. This can substantially reduce the received interference at the users. Higher chance of experiencing an NLOS channel is generally not problematic in this case as cooperative communication compensate for the decline of information-bearing signals. 2) RSHR has the highest EE regardless of popularity exponent. Importantly, it is able to increase the EE by up to 800% (resp. almost 400%) compared to LRU and Hit-rate schemes in high-rise (resp. sub-urban) environment, which is in our view impressive regarding its simplicity. 3) The EE under MPCP increases steadily with $\kappa$, which is inline with what we expect; the more skewed is the distribution of popularity, the higher chance that the most-popular contents is requested. Finally, 4) Under MPRC, the EE falls in between RSHR and MPCP schemes. In effect, when $\kappa > 1.3$, both schemes MPRC and MPCP result in the same EE. Note that for this regime the MPCP performs too closely to the EE of RSHR, which substantiate our previous discussions in Section IV-A.

*2) Impact of Cooperation Zone Radius $X_{\text{cop}}$:* Fig. 4 demonstrates the impact of $X_{\text{cop}}$ on EE for communication environments of high-rise and dense-urban. From this figures we observe several interesting trends: 1) Under MPCP the EE with respect to $X_{\text{cop}}$ follows an inverted-U curve; the EE reaches its pinnacle at $X_{\text{cop}}^* \approx 2$. Interestingly, for $X_{\text{cop}} \leq X_{\text{cop}}^*$, the MPCP scheme performs as good as RSHR and MPRC schemes. 2) When $X_{\text{cop}} > 2.2$, the EE under RSHR scheme keeps growing by the expansion of the cooperation zone although with the rate not as high as the observed growth rate for $X_{\text{cop}} < 2$. We note that when $X_{\text{cop}} = 4$ km, under RSHR the EE is doubled compared to the performance of MPCP. 3) We again observe that MPCP falls in between MPCP and RSHR. Comparing performance of MPCP and MPRC we conclude that when $X_{\text{cop}} \lesssim 3$, it is almost optimal to merely cache the most popular contents, but by the expansion of the cooperation zone it becomes more suitable to also cache less popular contents too. 4) We note that in high-rise environment the EE is much higher than that of the dense-urban environment, which again is attributable to the substantial reduction of LOS interference in the former case. Finally, 5) compared to Hit-rate and LRU, the MPCP (resp. MPRC) scheme improves the EE up to 800% (resp. 700%).

*3) Impact of Density $\lambda$:* In Fig. 5 we show the EE versus the density of transmitters. As seen, regardless of the environment choice as well as the caching policy the EE is bell-shaped where the optimal density rendering the maximization of EE, $\lambda^* \approx 0.1$, coincides in all considered content placement algorithms. The growth of EE for $\lambda < 0.1$ is attributable to the increase of signal strength as more UAVs find themselves in the cooperation zone carrying the requested contents. Nevertheless, increasing density further, i.e., $\lambda > 1$, causes a dramatic growth of the accumulated LOS interfering signals, cancelling out the harnessed signal strength by the cooperative communication. While adopting more sophisticated content placement strategies, e.g., RSHR and MPRC, can substantially increase the EE performance compared to Hit-rate and LRU schemes (here by more than 200%), they are basically impotent to forestall the crash of EE.

*Remark 2:* The low coverage probability of dense terrestrial communication networks has highlighted in the literature, see, e.g., [42–44]. The current literature of caching in terrestrial networks commonly ignores the impact of LOS/NLOS path-loss propagation and resort to the idealistic path-loss model. Although the focus of our study in this paper is on UAV communication, the result of Fig. 5 hints that the results of [15–19, 23, 24] may solely be applicable when the network is moderately densified. ∎

*4) Impact of Vertical Displacement:* Now we study the impact of vertical displacement on the EE. Results can be found in Fig. 6 that depicts the ratio $\frac{\eta^*}{\eta^*(H_0)}$ versus $H_0$. Here, quantity $\eta^*(H_0)$ is the EE performance of content placement problem for given altitude $H_0$—we refer to it as baseline EE performance in the following. In contrast, quantity $\eta^*$ takes advantage of possible vertical displacement. As seen from Fig. 6, under RSHR, MPRC, and MPCP strategies vertical displacement is able to substantially increases the EE compared to the baseline EE performance, where the harnessed performance growth increases steadily by increasing $H_0$. In effect, by increasing $H_0$ the baseline performance could suffer from two fronts: 1) the growth of the path-loss attenuation by height, and 2) the higher level of LOS interfering signals. Interestingly, from Fig. 6 we observe that neither of Hit-rate nor LRU schemes are able to absorb any significant EE growth against the baseline EE performance. This implies the importance of suitable content placement strategy in UAV communications. Finally, from Fig. 6 we see that RSHR scheme can in general improve the EE much better than MPRC as well as MPCP algorithms.

*5) Impact of Cache Size:* Now we investigate the impact of cache size on the EE in Fig. 7. Here we depict $\eta^*$ versus $S/F$. We make several observations from these illustrations. 1)

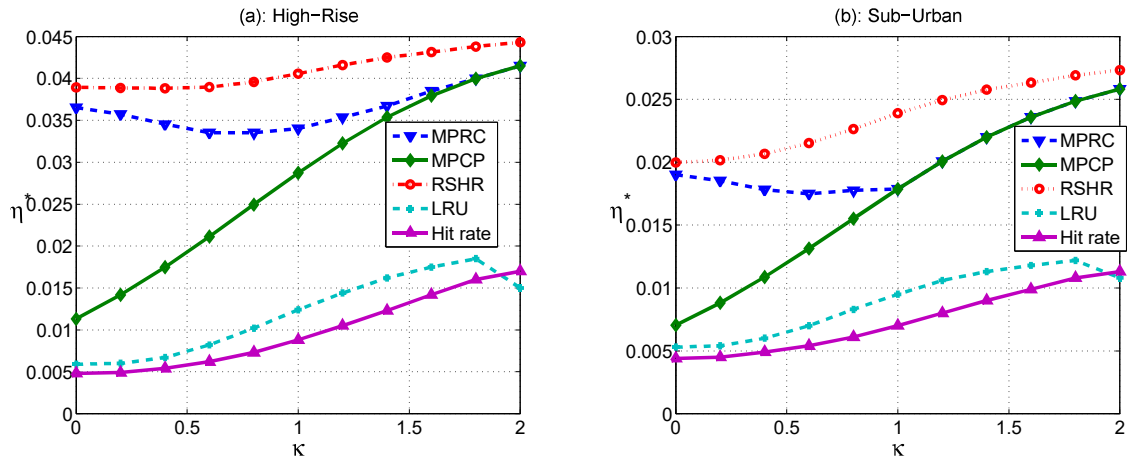

Fig. 3.  Energy-efficiency versus $\kappa$, where $X_{\text{cop}} = 2$, $\lambda = 10^{-2}$, $S = 5$, and $F = 20$.

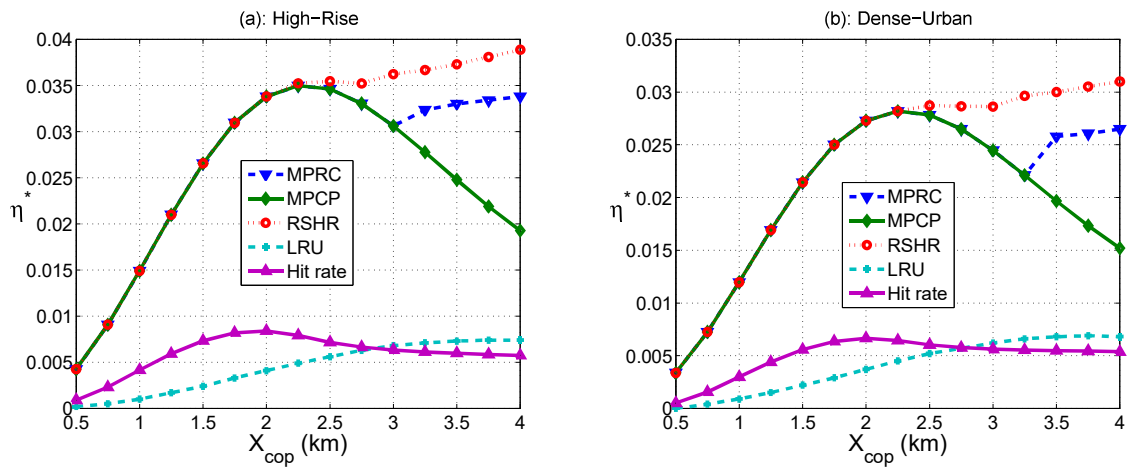

Fig. 4.  Energy-efficiency versus $X_{\text{cop}}$, where $\lambda = 10^{-1}$, $S = 5$, and $F = 20$.

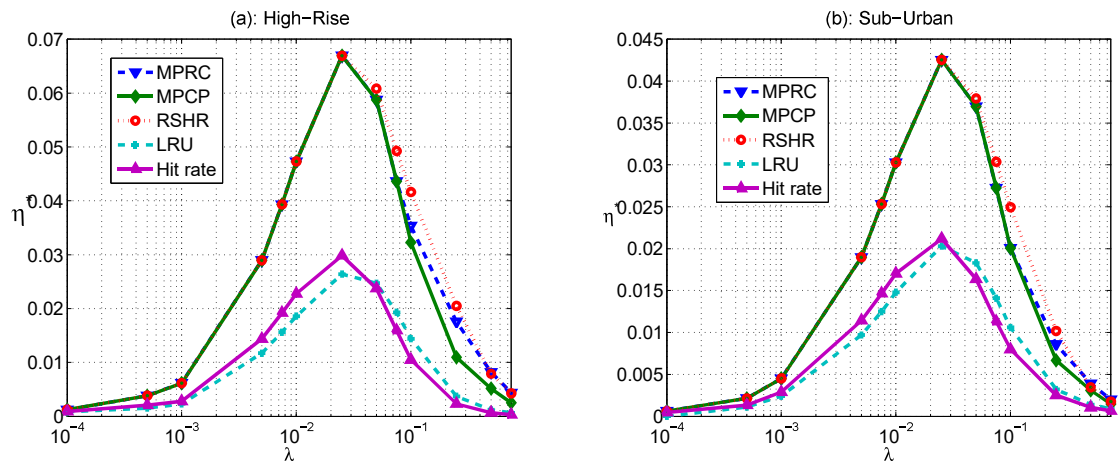

Fig. 5.  Energy-efficiency versus $\lambda$, where $F = 50$, $S = 10$, and $\kappa = 1.2$, and $X_{\text{cop}} = 2$.

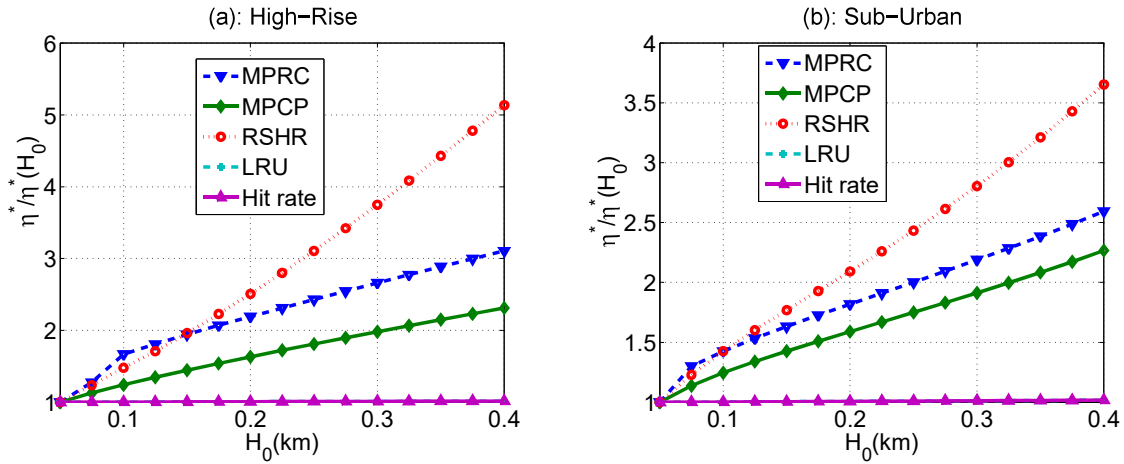

Fig. 6. $\frac{\eta^*}{\eta^*(H_0)}$ versus $H_0$, where $\lambda = 10^{-1}$, $F = 60$, $S = 15$, and $\kappa = 0.8$, and $X_{\text{cop}} = 3$.

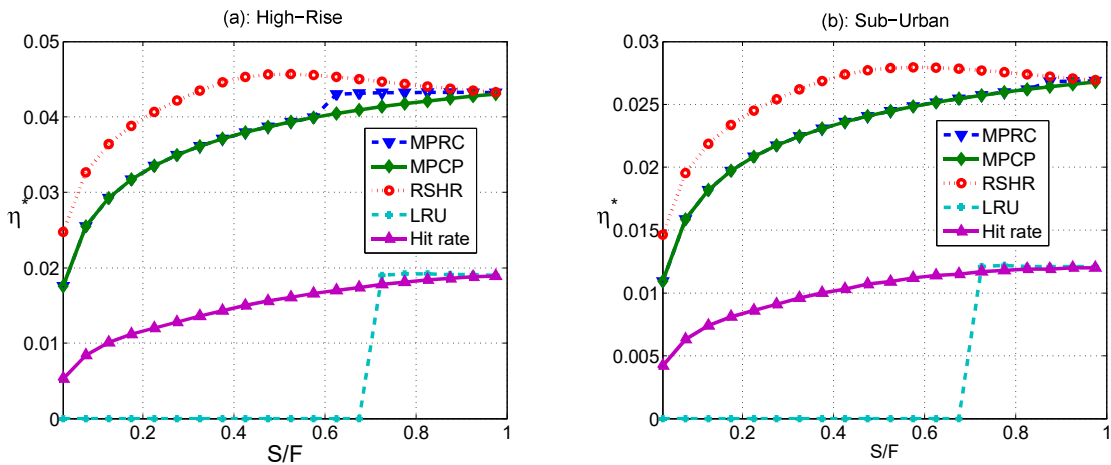

Fig. 7. Energy-efficiency versus $S/F$, where $\lambda = 10^{-1}$, $F = 200$, and $\kappa = 1$, and $X_{\text{cop}} = 4$.

Under RSHR there is an optimal cache size that results in the optimization of EE. In effect, only caching half of the library size is enough in order to achieve the maximum EE. None of the other schemes predict this phenomenon. Consequently, adopting inappropriate caching scheme may call for increasing the cache size. 2) In high-rise environment, MPRC slightly performs better than MPCP algorithm when $S/F > 0.5$. 3) When $S/F \leq 0.6$, the EE under LRU is almost zero. However, when $S/F \approx 0.6$ the EE under LRU jumps swiftly to the EE performance resulted from Hit-rate.

## VI. CONCLUSIONS AND REMARKS

We studied the EE in UAV-enabled Fog-RAN by devising an optimization problem to optimally place contents at the caches of UAVs and to choose the height of UAVs. We developed two new content placement strategies, namely RSHR and MPRC, with very light computational burdens. Our numerical results show that adopting these algorithms one is able to improve the EE performance substantially compared to the common approaches considered in the literature, e.g., LRU, the most-popular, and Hit-rate. We further demonstrated the significance of suitable caching in UAV-enabled Fog-RAN by showing that 1) while under LRU and Hit-rate schemes there is no benefit in vertically displacing UAVs, under our algorithms one is able to increase the EE performance by up to $600\%$; 2) under optimal caching scheme it is merely enough to cache half of the popular contents to reach the maximum EE; 3) there is the optimal density of UAVs that maximize the EE performance regardless of the height and content placement strategy; 4) via proper caching schemes one can increase the size of cooperation zone in order to steadily increase the EE performance, which is not the cases of LRU, the most-popular, and Hit-rate schemes.

## APPENDIX

### A. Proof of Proposition 1

Consider $X_{\min}$ as the closest UAV to the origin. Further, let us denote $N_c = |\tilde{\Phi}_c \backslash X_{\min}|$, which, excusing $X_{\min}$, is the number of UAVs with $c \in \mathcal{S}_X$ where $\|X_{\min}\| \leq \|X\| \leq X_{\text{cop}}$.

We then write $\gamma_c(H, p_c)$ as

$$\gamma_c(H, p_c) = \mathbb{E}_{\|X_{\min}\| \le X_{\text{cop}}} \sum_{n_c=0}^{\infty} \mathbb{P}\{N_c = n_c\} \gamma_c^{n_c}, \quad (21)$$

in which

$$\gamma_c^{n_c} = \frac{1}{n_c + 1} \mathbb{E}\Big[\log\Big(1 + \frac{P_r(x) + \sum_{m=1}^{n_c} P_r(X_m)}{I_c + \sigma^2}\Big)\Big], \quad (22)$$

where the expectation is over any randomness involved in the model conditioned on $x = \|X_{\min}\|$. Note that random variable $N_c$ is distributed according to Poisson distribution and admits the probability mass function (pmf)

$$\mathbb{P}\{N_c = n_c\} = \frac{(\pi \lambda p_c (X_{\text{cop}}^2 - x^2))^{n_c}}{n_c!} e^{-\pi \lambda p_c (X_{\text{cop}}^2 - x^2)}. \quad (23)$$

Substituting (22) and (23) into (21), we then write $\eta_c$ via the following integration formula

$$\gamma_c(H, p_c) = 2\pi \lambda p_c \int_0^{X_{\text{cop}}} x e^{-\pi \lambda p_c x^2} \Big[ e^{-\pi \lambda p_c (X_{\text{cop}}^2 - x^2)} \gamma_c^0$$

$$+ \sum_{n_c=1}^{\infty} \frac{(\pi \lambda p_c (X_{\text{cop}}^2 - x^2))^{n_c}}{n_c!} e^{-\pi \lambda p_c (X_{\text{cop}}^2 - x^2)} \gamma_c^{n_c} \Big] dx$$

$$= \frac{2\pi \lambda p_c}{e^{p_c \Theta_{\text{cop}}}} \int_0^{X_{\text{cop}}} x \Bigg( \underbrace{\mathbb{E}\Big[\log\Big(1 + \frac{P_r(x)}{I_c + \frac{\sigma^2}{P}}\Big)\Big]}_{T_1(x)} + \sum_{n_c=1}^{\infty} \frac{(\pi \lambda p_c)^{n_c}}{(n_c + 1)!}$$

$$(X_{\text{cop}}^2 - x^2))^{n_c} \underbrace{\mathbb{E}\Big[\log\Big(1 + \frac{P_r(x) + \sum_{m=1}^{n_c} P_r(X_m)}{I_c + \frac{\sigma^2}{P}}\Big)\Big]}_{T_2(x)} \Bigg) dx. \quad (24)$$

We now evaluate $T_2(x)$ observing that $T_1(x)$ is a specific case of $T_2(x)$ (setting $n_c = 0$). Using $\log(1 + x) = \int_0^{\infty} e^{-v}/v(1 - e^{-vx}) dv$, we firstly write

$$T_2(x) = \mathbb{E} \int_0^{\infty} \frac{e^{-v}}{v}\Big(1 - e^{-v\frac{P_r(x) + \sum_{m=1}^{n_c} P_r(X_m)}{I_c + \frac{\sigma^2}{P}}}\Big) dv$$

$$= \int_0^{\infty} \frac{\mathbb{E}_{I_c}[e^{-vI_c}]}{ve^{v\frac{\sigma^2}{P}}} \Big(1 - \mathbb{E}\big[e^{-vP_r(x)} e^{-v\sum_{m=1}^{n_c} P_r(X_m)}\big]\Big) dv$$

$$= \int_0^{\infty} \frac{\mathcal{L}_{I_c}(v)}{ve^{v\frac{\sigma^2}{P}}} \Big(1 - \underbrace{\mathbb{E}[e^{-vP_r(x)}]}_{\mathcal{L}_1(x,v)} \underbrace{\mathbb{E}\big[e^{-v\sum_{m=1}^{n_c} P_r(X_m)}\big]}_{\mathcal{L}_2(x,v)}\Big) dv. \quad (25)$$

Straightforward manipulations imply that $\mathcal{L}_1(x, v)$, the Laplace transform of the signal power from the closest UAV, is

$$\mathcal{L}_1(x, v) = \sum_{l_x \in \{L, N\}} p_{l_x}(x) \mathbb{E}_{V_x^{l_x}} \mathbb{E}_{W_x^{l_x}} e^{-xL_{l_x}(x) V_x^{l_x} W_x^{l_x}}$$

$$= \sum_{l_x \in \{L, N\}} \int_0^{\infty} \frac{p_{l_x}(x)}{\sqrt{(2\pi \sigma_x^{l_x})^2}} e^{-\frac{(u - \mu^{l_x})^2}{2(\sigma_x^{l_x})^2}} \frac{1}{\big(1 + v\frac{L_{l_x}(x) 10^{u/10}}{\overline{W}_{l_x}}\big)^{\overline{W}_{l_x}}} du. \quad (26)$$

On the other hand, $\mathcal{L}_2(x, v)$, the Laplace transform of aggregate signal powers from helpers in the cooperation zone, can also be calculated as the following

$$\mathcal{L}_2(x, v) = \int_x^{X_{\text{cop}}} \cdots \int_x^{X_{\text{cop}}} \prod_{m=1}^{n_c} \frac{2x_m}{X_{\text{cop}}^2 - x^2} \mathbb{E} e^{-vP_r(x_m)} dx_m$$

$$= \prod_{m=1}^{n_c} \int_x^{X_{\text{cop}}} \frac{2x_m \mathcal{L}_1(x_m, v)}{X_{\text{cop}}^2 - x^2} dx_m = \Big(\int_x^{X_{\text{cop}}} \frac{2y \mathcal{L}_1(y, v)}{X_{\text{cop}}^2 - x^2} dy\Big)^{n_c}$$

$$= \frac{(\Psi(x, v))^{n_c}}{(X_{\text{cop}}^2 - x^2)^{n_c}}. \quad (27)$$

In (25) $\mathcal{L}_{I_c}(v)$ is the Laplace transform of the interference. As the interference is originated from two independent sources—UAVs not having file $f_c$ in their caches and the UAVs holding a copy of file $f_c$ in their caches but are flying outside cooperation zone—we can evaluate $\mathcal{L}_{I_c}(v)$ as the following

$$\mathcal{L}_{I_c}(v) = \underbrace{\mathbb{E}_{\Phi \setminus \Phi_c, \{P_r(Z)\}_{Z \in \Phi \setminus \Phi_c}}\Big[e^{-\sum_{Z \in \Phi \setminus \Phi_c} P_r(Z)}\Big]}_{T_3}$$

$$\underbrace{\mathbb{E}_{\Phi_c \setminus \bar{\Phi}_c, \{P_r(Y)\}_{Y \in \Phi_c \setminus \bar{\Phi}_c}}\Big[e^{-\sum_{Y \in \Phi_c \setminus \bar{\Phi}_c} vP_r(Y)}\Big]}_{T_4} dv. \quad (28)$$

We evaluate $T_3$ and $T_4$ in the following. For $T_3$, we note that 1) each communication link independently undergoes LOS mode, 2) shadowing and fading power gains on each communication link are independent, and 3) shadowing (fading) power gains across communication gains are independent. Therefore,

$$T_3 = \prod_{c' \ne c} \mathbb{E}_{\Phi_{c'}, \{P_r(Z), \|Z\| \le X_{\text{cop}}\}} e^{-\sum_{Z \in \Phi_{c'}} vP_r(Z)}$$

$$= \prod_{c' \ne c} \mathbb{E}_{\Phi_{c'}, \|Z\| \le X_{\text{cop}}} \prod_{Z \in \Phi_{c'}} \mathbb{E}_{\{L(Z), V_Z, W_Z\}} e^{-vL(Z)V_Z W_Z}$$

$$= \prod_{c' \ne c} \mathbb{E}_{\Phi_{c'}, \|Z\| \le X_{\text{cop}}} \prod_{Z \in \Phi_{c'}} \mathcal{L}_1(Z, v)$$

$$= \prod_{c' \ne c} e^{-2\pi p_{c'} \lambda \int_0^{X_{\text{cop}}} (1 - \mathcal{L}_1(z, v)) dz}$$

$$= e^{-2\pi (1 - p_c) \lambda \int_0^{X_{\text{cop}}} (1 - \mathcal{L}_1(z, v)) dz}, \quad (29)$$

where we also insert normal distribution with distance-dependent variance as in (5) and apply Laplace generation functional of HPPP as in [10]. To evaluate $T_3$ we follow the same line of argument above that yields

$$T_4 = e^{-2\pi p_c \lambda \int_{X_{\text{cop}}}^{\infty} (1 - \mathcal{L}_1(z, v)) dz} \quad (30)$$

Consequently, $T_2(x)$ is formulated as

$$T_2(x) = \int\limits_0^\infty \frac{e^{-v\frac{\sigma^2}{P}}}{v} \mathcal{L}_{I_c}(v)\big(1 - \mathcal{L}_1(x,v)(\Psi(x,v))^{n_c}\big)dv.$$

Likewise, one can straightforwardly evaluate $T_1(x)$ as

$$T_1(x) = \int\limits_0^\infty \frac{e^{-v\frac{\sigma^2}{P}}}{v} \mathcal{L}_{I_c}(v)(1 - \mathcal{L}_1(x,v))dv$$

Thus,

$$\gamma_c(H,p_c) = \frac{2\pi\lambda p_c}{e^{p_c\Theta_{\text{cop}}}} \int\limits_0^{X_{\text{cop}}} \int\limits_0^\infty \frac{\mathcal{L}_{I_c}(v)}{ve^{v\frac{\sigma^2}{P}}} x \bigg(1 - \mathcal{L}_1(x,v)$$

$$+ \sum_{n_c=1}^\infty \frac{(\pi\lambda p_c(X_{\text{cop}}^2 - x^2))^{n_c}}{(n_c+1)!}\big(1 - \mathcal{L}_1(x,v)(\Psi(x,v))^{n_c}\big)\bigg)dvdx$$

$$= \frac{2\pi\lambda p_c}{e^{p_c\Theta_{\text{cop}}}} \int\limits_0^{X_{\text{cop}}} \int\limits_0^\infty \frac{\mathcal{L}_{I_c}(v)}{ve^{v\frac{\sigma^2}{P}}} x \bigg(1 + \frac{\sum_{n_c=1}^\infty \frac{(\pi\lambda p_c(X_{\text{cop}}^2 - x^2))^{n_c+1}}{(n_c+1)!}}{\pi\lambda p_c(X_{\text{cop}}^2 - x^2)}$$

$$- \mathcal{L}_1(x,v) - \frac{\mathcal{L}_1(x,v)}{\pi\lambda p_c\Psi(x,v)} \sum_{n_c=1}^\infty \frac{(\pi\lambda p_c\Psi(x,v)^{n_c+1})}{(n_c+1)!}\bigg)dvdx$$

$$= \frac{2\pi\lambda p_c}{e^{p_c\Theta_{\text{cop}}}} \int\limits_0^{X_{\text{cop}}} \int\limits_0^\infty \frac{\mathcal{L}_{I_c}(v)}{ve^{v\frac{\sigma^2}{P}}} x \bigg(\frac{e^{\pi\lambda p_c(X_{\text{cop}}^2 - x^2)} - 1}{\pi\lambda p_c(X_{\text{cop}}^2 - x^2)}$$

$$- \mathcal{L}_1(x,v)\frac{e^{\pi\lambda p_c\Psi(x,v)} - 1}{\pi\lambda p_c\Psi(x,v)}\bigg)dvdx,$$

which proves the proposition.

### B. Proof of Proposition 2

We start by approximating the received information-bearing signal powers via $\sum_{X\in\tilde{\Phi}_c} P_r(X) \approx P_r(X_{\min}) + |\tilde{\Phi}_c - 1|P_r(X'_{\text{cop}})$, where $X'_{\text{cop}}$ is a point with distance $0.5\|X_{\min}\| + 0.5X_{\text{cop}}$ from the origin. In effect, we assume that all $|\tilde{\Phi}_c - 1|$ UAVs in the cooperation zone—excluding the closest one to the origin—are located in the 2-D distance half the way between $X_{\min}$ and the border of the cooperation zone. Therefore, we approximate the SINR as the following

$$\text{SINR}_c \approx \hat{\text{SINR}}_c = \frac{P_r(X_{\min}) + |\tilde{\Phi}_c - 1|P_r(X'_{\text{cop}})}{\sigma^2/P + I_c}, \quad (31)$$

which implies that the achievable capacity under the request of file $f_c$ can be approximated as

$$\hat{R}_c(n_c) = \mathbb{E}\left[\log\left(1 + \hat{\text{SINR}}_c\right)||\tilde{\Phi}_c| = n_c\right]$$

$$= \mathbb{E}\left[\log\left(1 + \frac{P_r(X_{\min}) + (n_c-1)P_r(X'_{\text{cop}})}{\sigma^2/P + I_c}\right)\right]$$

$$\leq \log\left(1 + \mathbb{E}[(P_r(X_{\min}) + (n_c-1)P_r(X'_{\text{cop}}))]\mathbb{E}[\frac{1}{\sigma^2/P + I_c}]\right),$$

where the upper-bound is due to Jensen's inequality and the independence of the nominator and denominator of (31). Using the fact that $\mathbb{E}[\frac{1}{\sigma^2/P + I_c}] = \int\limits_0^\infty e^{-v\sigma^2/P}\mathcal{L}_{I_c}(v)dv$, we then have

$$\hat{R}_c(n_c) \approx \log\bigg(1 + 2\pi\lambda p_c \int\limits_0^{X_{\text{cop}}} xe^{-\pi\lambda p_c x^2}\big(\overline{P}_r(x)$$

$$+ (n_c-1)\overline{P}_r(0.5x + 0.5X_{\text{cop}})\big)dx \int\limits_0^\infty \frac{\mathcal{L}_{I_c}(v)}{e^{\frac{\sigma^2}{P}v}}dv\bigg). \quad (32)$$

By plugging (32) into

$$\gamma_c(H,p_c) \approx \sum_{n_c=1}^\infty e^{-\Theta_{\text{cop}}p_c}\frac{(\Theta_{\text{cop}}p_c)^{n_c}}{n_c!}\frac{\hat{R}_c(n_c)}{n_c},$$

that proves the proposition.